\documentclass{ws-ijmpcs}
\usepackage{bbm}
\usepackage{subfigure}
\usepackage{multirow}
\usepackage{hyperref}

\begin{document}

\markboth{W.C. Chang}
{Unpolarized Drell-Yan program of COMPASS experiment}

%
%

\title{Nucleon partonic spin structure to be explored by the
  unpolarized Drell-Yan program of COMPASS experiment at CERN}

\author{Wen-Chen Chang\footnote{on behalf of COMPASS Collaboration}}

\address{Institute of Physics, Academia Sinica, Taipei 11529, Taiwan \\
changwc@phys.sinica.edu.tw}

\maketitle

\begin{history}
\received{Day Month Year}
\revised{Day Month Year}
\published{Day Month Year}
\end{history}

\begin{abstract}

The observation of the violation of Lam-Tung relation in the $\pi N$
Drell-Yan process triggered many theoretical speculations. The TMD
Boer-Mulders functions characterizing the correlation of transverse
momentum and transverse spin for partons in unpolarized hadrons could
nicely account for the violation. The COMPASS experiment at CERN will
measure the angular distributions of dimuons from the unpolarized
Drell-Yan process over a wide kinematic region and study the beam
particle dependence. Significant statistics is expected from a
successful run in 2015 which will bring further understanding of the
origin of the violation of Lam-Tung relation and of the partonic
transverse spin structure of the nucleon.


\end{abstract}

\ccode{PACS numbers:}

\section{Introduction}
\label{sec:intro}

The continuum of dilepton from hadron-hadron collisions has been
successfully described by the Drell-Yan parton model\cite{drell-yan}
where collinear quarks and antiquarks ($q\bar{q}$) from two individual
hadrons electromagnetically annihilate into a virtual photon. The
predictions of the scaling behavior, the $A$-dependence of production
cross section, and the polar angle distribution of leptons have been
experimentally verified\cite{dyreview1,dyreview2}. Furthermore the
quantitative description of the transverse momentum distribution of
lepton pairs and the enhancement of the cross section over the naive
Drell-Yan estimate do call for the perturbative QCD effect of gluon
emission and absorption by the partons.

Drell-Yan process has become one of the most intensively studied
processes in QCD\cite{dyreview3,dyreview4} and an effective tool to
learn about the partonic structure of hadrons, ranging from the parton
distribution functions (PDFs) to transverse momentum dependent (TMD)
parton density distributions. Nonetheless there remain challenges to
understand why Lam-Tung relation failed in the dilepton azimuthal
angular distributions of the $\pi N$ Drell-Yan
reaction\cite{dyreview5}.

Since 2002 the COMPASS collaboration at CERN has been one of the major
experiments for studying both unpolarized and polarized parton
structures of nucleons, and TMD
distributions\cite{COMPASS_talk1}. Nonzero TMD Sivers functions for
the valence quarks of nucleons have been measured in semi-inclusive
DIS (SIDIS)\cite{COMPASS_talk2}. In the second phase of the
experiment\cite{COMPASSII}, COMPASS will seek for the Sivers functions
in the first-ever polarized Drell-Yan experiment with a
transversely-polarized NH$_3$ target\cite{COMPASS_talk3}. This
measurement will be an important experimental test of the
long-predicted sign change of Sivers functions in polarized SIDIS and
Drell-Yan processes.

In this talk we report the impact on TMD spin physics to be brought by
the simultaneous measurement of the unpolarized Drell-Yan process in
COMPASS. The violation of Lam-Tung relation in the $\pi N$ Drell-Yan
process and the interpretation of this phenomenon by TMD Boer-Mulders
functions and other theoretical models will be introduced. A
comprehensive measurement in the coming unpolarized Drell-Yan program
of CERN/COMPASS experiment shall shed light on this problem.

\section{Angular Distributions of Unpolarized Drell-Yan process}
\label{sec:angdist}

Assuming dominance of the single-photon process, the angular
distribution of leptons from the unpolarized Drell-Yan process could
be expressed by ${d \sigma}/{d \Omega} \propto (1+ \lambda
\cos^2\theta + \mu \sin2\theta \cos \phi + \frac{\nu}{2}\sin^2\theta
\cos2\phi)$, where $\theta$ and $\phi$ are the polar and azimuthal
angles of the decay leptons in the virtual photon rest frame, and
$\lambda$, $\mu$ and $\nu$ are angular parameters. The virtual photon
from collinear $q\bar{q}$ annihilation is transversely polarized and
thus the decay angular distribution of muons is proportional to
$(1+\cos^2\theta)$, i.e. $\lambda =1$ and $\mu=\nu=0$. The early
experimental measurement agreed with this prediction.

In 1978, Lam and Tung studied the quark intrinsic transverse momentum
effects on the angular distributions of lepton
pairs\cite{LamTung}. They found that QCD effect could lead to $\lambda
\neq 1$ and $\mu,\nu \neq 0$ but the relation $1-\lambda=2\mu$, the
so-called ``Lam-Tung relation'' holds for NLO QCD effect.  Therefore
this relation provides a unique opportunity to test the ``QCD-improved
quark-parton model''\cite{LamTung}.

\begin{table}[hbt]
\tbl{Theoretical interpretations of Lam-Tung violation in the $\pi N$
  Drell-Yan process.}
{\begin{tabular}{@{}cccc@{}} \toprule
& Boer-Mulders & QCD & Glauber gluon \\
& Function  & chromo-magnetic effect \\ \colrule
origin of effect & hadron & QCD vacuum & pion specific \\
quark flavor dependence & yes & no & no \\
beam dependence & yes & no & yes \\
large $p_T$ limit & 0 & finite & 0 \\ \botrule
\end{tabular} \label{ta1}}
\end{table}

The very first measurement by the NA3\cite{NA3} experiment showed that
$\nu$ increases strongly toward large transverse momentum of the
lepton pair ($p_{T}$) but the Lam-Tung relation was roughly
preserved. Nevertheless the following measurements with better
statistics by the NA10\cite{NA10} and the E615\cite{E615} experiments
clearly identified a strong violation of Lam-Tung relation in the $\pi
N$ Drell-Yan process with nuclei and deuterium targets. The degree of
violation got stronger for dimuon pairs with large $p_T$.

The nonzero $\nu$ or $\cos 2\phi$ asymmetry comes from the
interference of virtual photon amplitudes with opposite
helicities\cite{Nachtmann,Boer05}. The occurrence of `` Lam-Tung
violation'' suggests certain mechanisms leading to the correlation of
the helicities of quark and anti-quark from the two individual
hadrons. Such an effect is commonly expected to be of nonperturbative
QCD nature.

Brandenburg {\it et al.}\cite{Nachtmann} proposed a factorization
breaking effect caused by nontrivial QCD vacuum where the
chromo-magnetic Sokolov-Ternov effect introduced a spin correlation
between the annihilating quark and anti-quark\cite{Nachtmann2}. On the
other hand, Boer\cite{Boer:LT} considered a hadronic effect of the
spin-orbit correlation of transversely polarized noncollinear partons
inside an unpolarized hadron,``Boer-Mulders
functions''($h_1^\bot$)\cite{BM}, and the angular parameter $\nu
\propto h_1^\bot(q_N)h_1^\bot(\bar{q}_{\pi})$. Recently it was shown
that the $p_T$ dependence of $\nu$ measured by NA10 and E615
experiments could be reasonably described by the Boer-Mulders
functions and the parton density of pions and nucleons constructed
within the light-front constituent model\cite{Pasquini:2014ppa}.

Table~\ref{ta1} lists the predictions given by the various theoretical
models including one considering Glauber gluons in the $k_T$
factorization theorem\cite{GlauberGluon}. It is clear that a
comprehensive measurement of dilepton angular distributions with
different types of beam like $\pi^\pm$, $p$, $K^\pm$ and $\bar p$ over
a wide kinematic range is required to clarify the mechanisms. For
example, the lack of Lam-Tung violation in the $pN$ Drell-Yan
process\cite{E866:pd} could be explained by a small Boer-Mulders
function for sea quarks in the target nucleon but might be less
compatible with the supposedly flavor-blind QCD vacuum effect.

\section{Unpolarized Drell-Yan Program of COMPASS Experiment at CERN}

\begin{figure*}[htbp]
\centering
\includegraphics[width=0.9\textwidth]{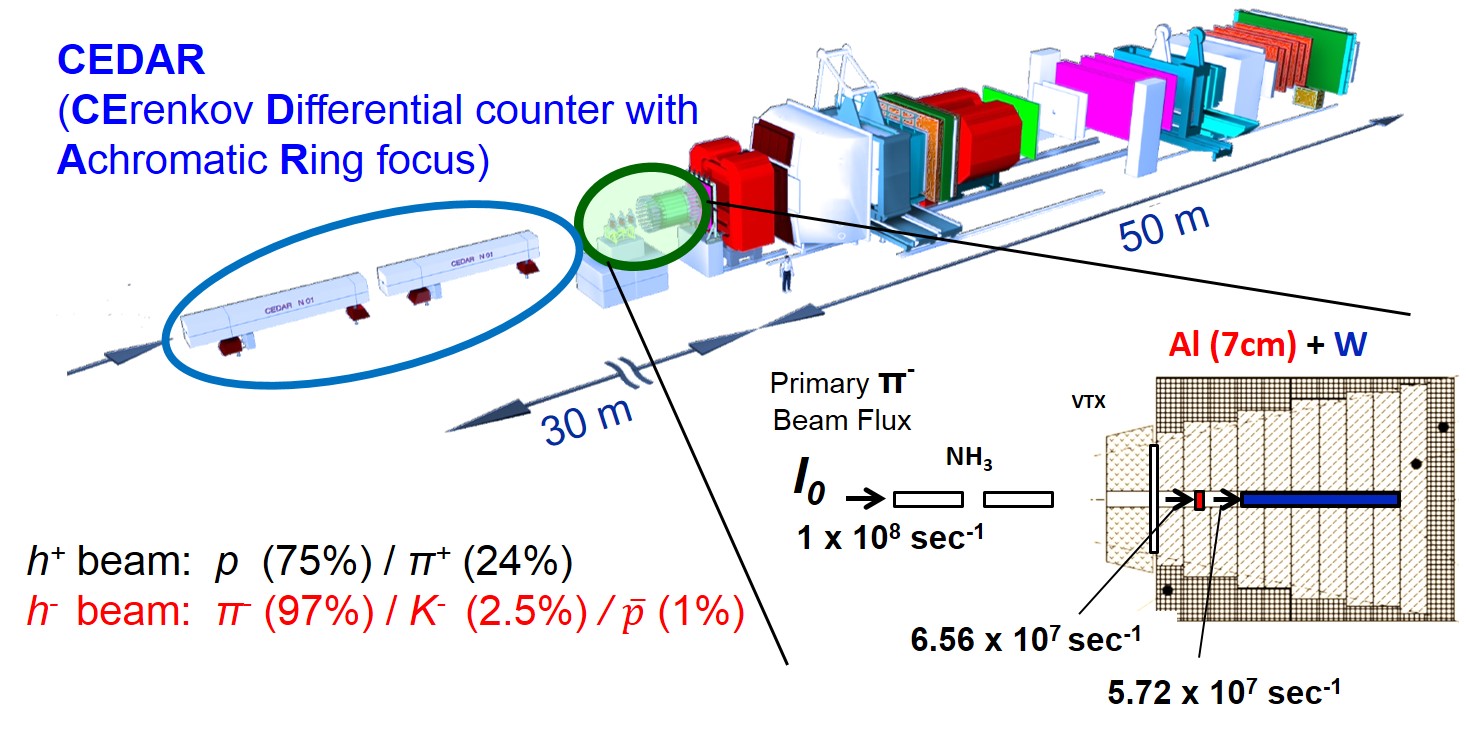}
\vspace*{-8pt}
\caption
[\protect{}]{The beam composition, detector setup and hadron absorber
  of COMPASS Drell-Yan experiment.}
\label{fig:setup}
\end{figure*}

As illustrated in Fig.~\ref{fig:setup}, the COMPASS experiment is
located at the SPS M2 secondary beam line at CERN. The secondary beam
produced by the slow-extracted 400 GeV/c primary proton beam can be
either muon or hadron secondary beams in a momentum range of 50 GeV/c
to 280 GeV/c. For 190-GeV momentum, the negatively-charged hadron beam
is composed of 97\% $\pi^-$, 2.5\% $K^-$ and 1\% $\bar p$ while it is
75\% $p$ and 24\% $\pi^+$ for the positively-charged beam. The flux of
the $\pi^-$ beam is $10^8$/s. There is a Cerenkov differential counter
(CEDAR) in front of the target which can provide particle
identification of incident beam. The CEDAR detector will enable the
measurement of kaon and antiproton-induced Drell-Yan processes in
COMPASS.

In order to detect the dimuon pairs from high-energy Drell-Yan
processes, a hadron absorber is required to prevent copious amounts of
produced hadrons from entering the spectrometer. The COMPASS hadron
absorber was made of aluminum and alumina (Al$_2$O$_3$) with a
downstream 20 cm layer of steel. It was designed to optimize both the
stopping power for hadron and the multiple scattering effect on
muons. Inside the absorber, a long tungsten (W) beam plug and a 7 cm
aluminum (Al) plate are placed. Both serve as nuclear targets for the
unpolarized Drell-Yan process.
 
The detector system is composed of a two-stage spectrometer such that
low-momentum charged particles in sideward direction as well as
high-momentum particles in forward direction could be well accepted.
Therefore COMPASS supersedes the past E615 experiment by one order of
magnitude in the overall acceptance of 40\%, especially that at large
$p_T$ region where the Boer-Mulders functions are more visible.

\begin{figure*}[htbp]
\begin{center}
\centering
\vspace*{8pt}
\subfigure[]
{\includegraphics[width=0.45\textwidth]{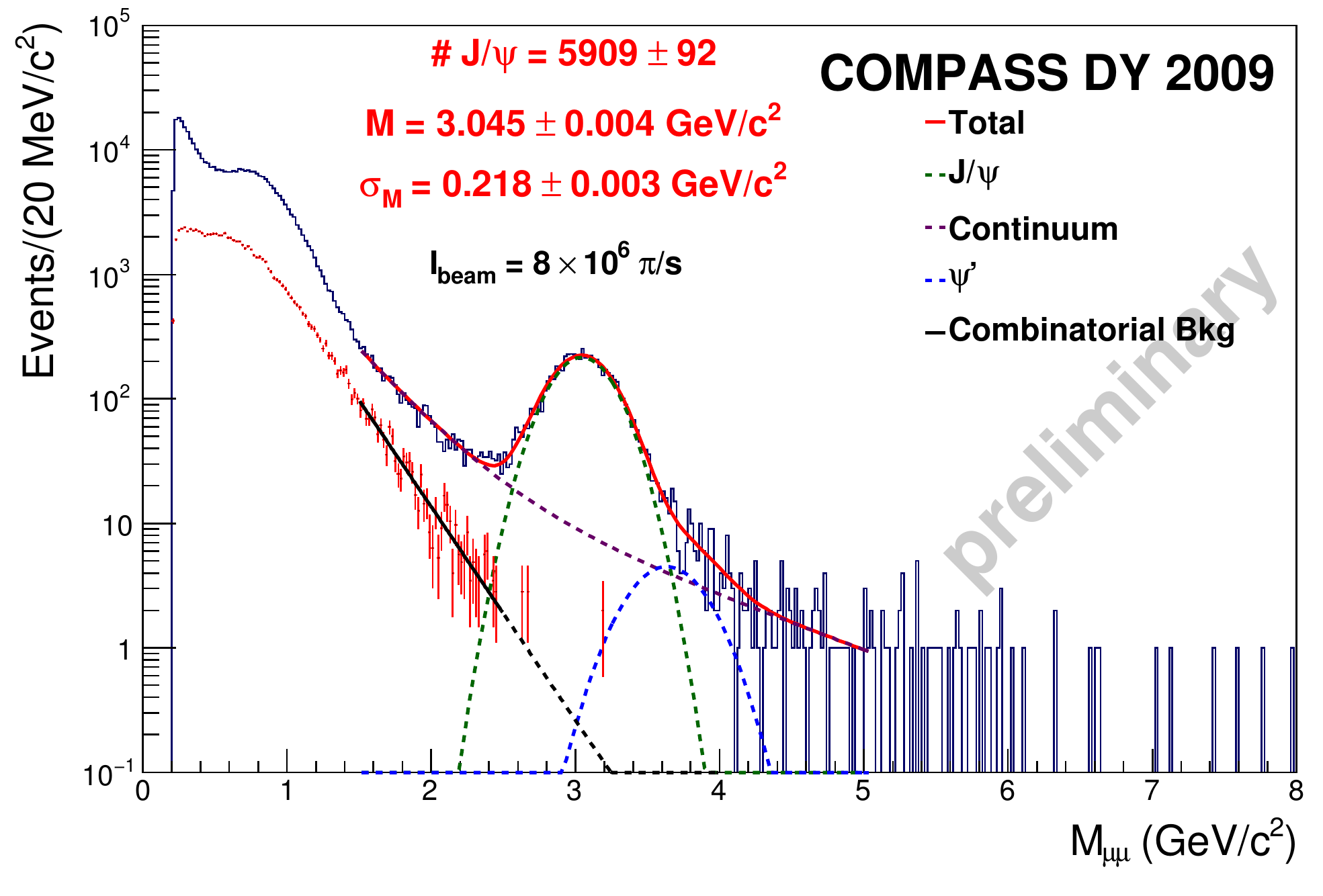}
\label{fig:dimuon1}}
\subfigure[]
{\includegraphics[width=0.45\textwidth]{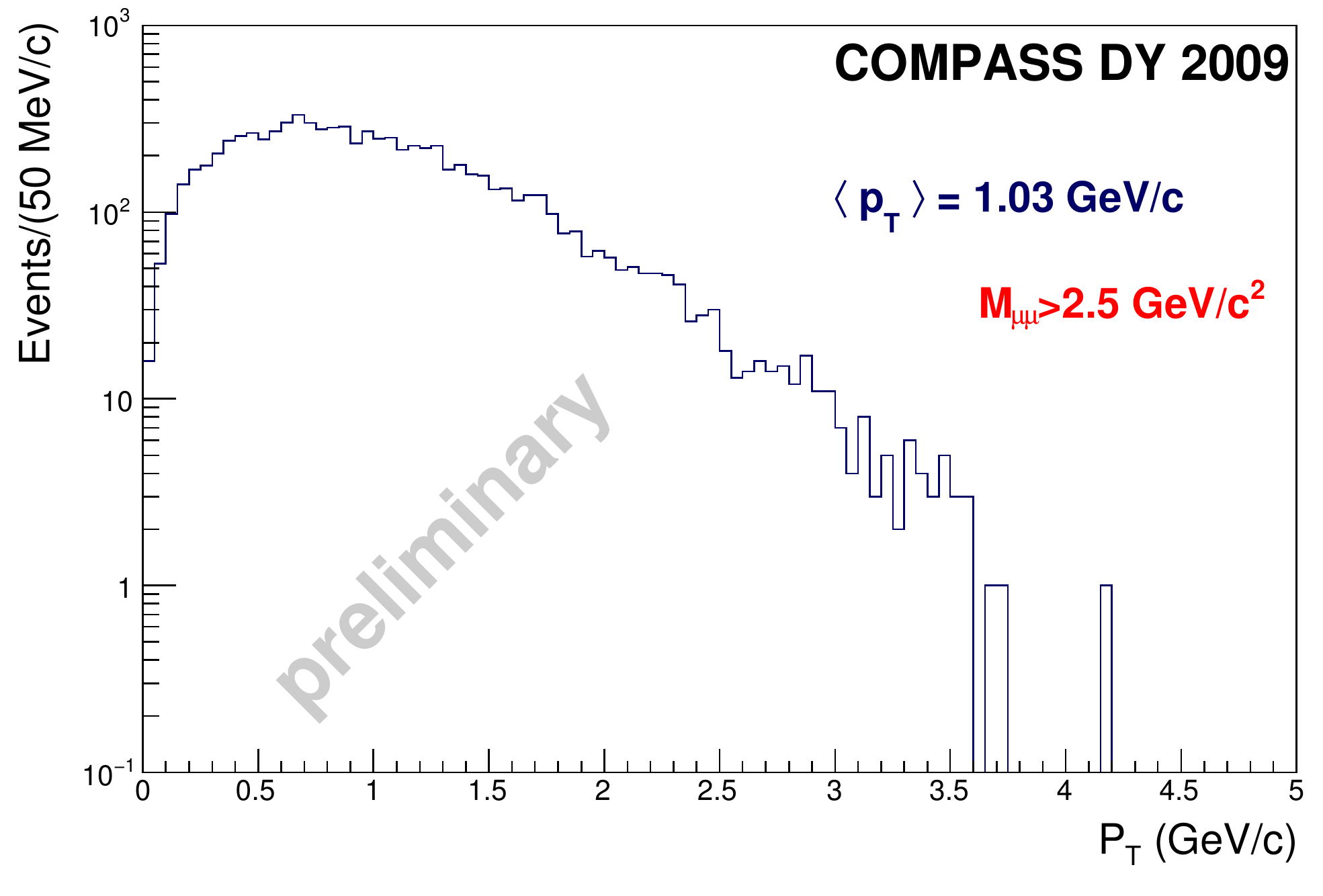}
\label{fig:dimuon2}}
\caption
[\protect{}] {(a) Invariant mass and (b) transverse momentum
  distributions of dimuon events from COMPASS 2009 test run.}
\label{fig:dimuon}
\end{center}
\end{figure*}

The feasibility of Drell-Yan measurement in COMPASS has been checked
by several test runs. In 2009 there was a 3-day test run where two
40-cm polyethylene cylinders and a prototype hadron absorber were
implemented. About 6000 $J/\psi$ were reconstructed from the data as
seen in Fig.~\ref{fig:dimuon}. The measured yield agreed nicely with
the one expected from Monte-Carlo simulation. The high-mass Drell-Yan
continuum was clearly observed.

\begin{figure*}[htbp]
\centering
\includegraphics[width=\textwidth]{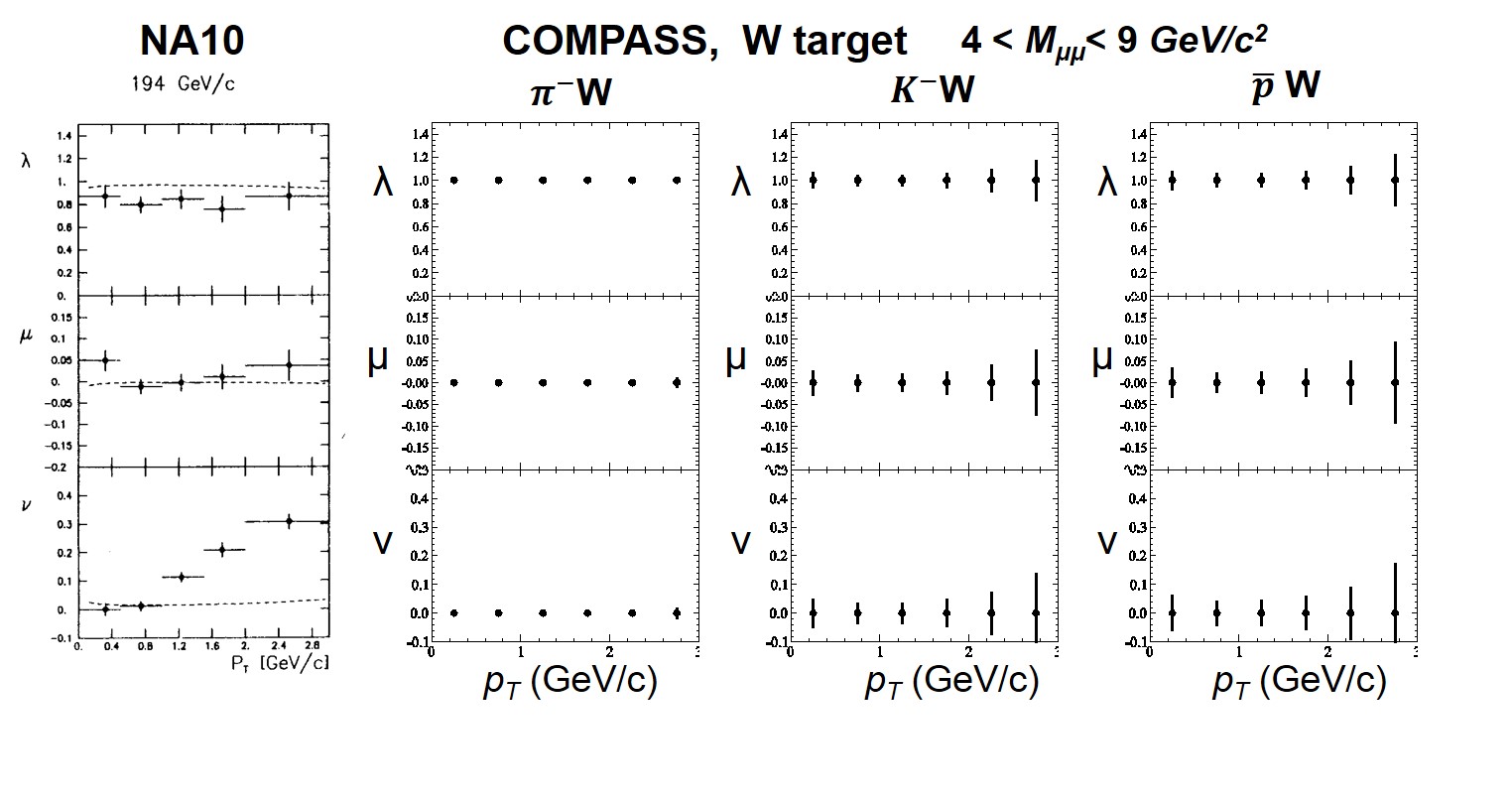}
\vspace*{-18pt}
\caption
[\protect{}]{Expected sensitivity of the angular parameters $\lambda$,
  $\mu$ and $\nu$ as a function of $p_T$ for $\pi^-$, $K^-$ and $\bar
  p$ beam with W target in COMPASS, compared with NA10's results for
  $\pi^-$ beam\cite{NA10}.}
\label{fig:angular_pt}
\end{figure*}

Assuming a beam flux of $10^8$/s for the $\pi^-$ beam, the expected
statistics of unpolarized high-mass ($4<M_{\mu\mu}<9 \rm{GeV}$)
Drell-Yan events from the three targets (NH$_3$, Al and W) in a
140-day beam time are 800k, 10k and 6k for $\pi$, $K$ and $\bar
p$-induced Drell-Yan events respectively. Compared to the statistics
from NA3\cite{NA3}, NA10\cite{NA10}, E537\cite{E537} and
E615\cite{E615}, a successful Drell-Yan experiment in 2015 will
increase the existing statistics of $\pi$, $K$ and $\bar p$-induced
Drell-Yan by a factor of 10. Fig.~\ref{fig:angular_pt} shows the
expected sensitivity of the angular parameter $\nu$ as a function of
$p_T$ for $\pi$, $K$ and $\bar p$-induced Drell-Yan events from W
target in COMPASS. Compared to NA10, the statistical error for
Drell-Yan data with $\pi^-$ beam will be significantly reduced and a
reasonable sensitivity could be achieved for data with $K^-$ and $\bar
p$ beam.

Commissioning of COMPASS Drell-Yan experiment was already completed
during mid-October - December, 2014. The physics run will take place
in year 2015. The beam time in 2016 and 2017 will be reserved for DVCS
program. Supposing that the LHC RUN2 is extended, there will be
possibility of one extra year for polarized (and unpolarized)
Drell-Yan run in 2018. For the long-term plan for Drell-Yan program in
COMPASS, polarized $^6$LiD and long LH$_2$ targets and high intensity
RF-separated antiproton/kaon beam is considered.

\section{Summary}

The TMD Boer-Mulders functions characterizing the spin-orbit
correlation of transversely polarized partons offer a reasonable
interpretation of the violation of Lam-Tung relation observed in the
$\pi N$ DY process. The COMPASS experiment will study not only the
Sivers functions in the polarized Drell-Yan process but also the
Boer-Mulders functions in the unpolarized DY process using $\pi$, $K$
and $\bar p$ beams over a wide kinematic range. A successful run in
2015 will hopefully bring further understanding of the origin of
Lam-Tung violation and the partonic structures of protons and pions.


\end{document}